\documentstyle[12pt,epsfig]{article}
\title{\hspace{9cm} {\large Budker INP 97--71} \\
\vspace{2.cm} 
 {\bf  High Energy  Photon-Photon Colliders 
 \footnote{ Talk at XI International Vavilov's Conference
on nonlinear optics  Novosibirsk, June 24-28, 1997}}}  
\author{Valery Telnov\footnote{Email: telnov@inp.nsk.su} \\
{\small\it Institute of Nuclear Physics,
630090, Novosibirsk, Russia}} 
\date{}                 
\begin{document}
\newcommand{\EP}{\mbox{e$^+$}}
\newcommand{\EM}{\mbox{e$^-$}}
\newcommand{\EPEM}{\mbox{e$^+$e$^-$}}
\newcommand{\EMEM}{\mbox{e$^-$e$^-$}}
\newcommand{\GG}{\mbox{$\gamma\gamma$}}
\newcommand{\GE}{\mbox{$\gamma$e}}
\newcommand{\GP}{\mbox{$\gamma$e$^+$}}
\newcommand{\TEV}{\mbox{TeV}}
\newcommand{\GEV}{\mbox{GeV}}
\newcommand{\LGG}{\mbox{$L_{\gamma\gamma}$}}
\newcommand{\EV}{\mbox{eV}}
\newcommand{\CM}{\mbox{cm}}
\newcommand{\MM}{\mbox{mm}}
\newcommand{\NM}{\mbox{nm}}
\newcommand{\MKM}{\mbox{$\mu$m}}
\newcommand{\SEC}{\mbox{s}}
\newcommand{\CMS}{\mbox{cm$^{-2}$s$^{-1}$}}
\newcommand{\MRAD}{\mbox{mrad}}
\newcommand{\IND}{\hspace*{\parindent}}
\newcommand{\E}{\mbox{$\epsilon$}}
\newcommand{\EN}{\mbox{$\epsilon_n$}}
\newcommand{\EI}{\mbox{$\epsilon_i$}}
\newcommand{\ENI}{\mbox{$\epsilon_{ni}$}}
\newcommand{\ENX}{\mbox{$\epsilon_{nx}$}}
\newcommand{\ENY}{\mbox{$\epsilon_{ny}$}}
\newcommand{\EX}{\mbox{$\epsilon_x$}}
\newcommand{\EY}{\mbox{$\epsilon_y$}}
\newcommand{\BI}{\mbox{$\beta_i$}}
\newcommand{\BX}{\mbox{$\beta_x$}}
\newcommand{\BY}{\mbox{$\beta_y$}}
\newcommand{\SX}{\mbox{$\sigma_x$}}
\newcommand{\SY}{\mbox{$\sigma_y$}}
\newcommand{\SZ}{\mbox{$\sigma_z$}}
\newcommand{\SI}{\mbox{$\sigma_i$}}
\newcommand{\SIP}{\mbox{$\sigma_i^{\prime}$}}
\maketitle
\begin{abstract}
 Using the laser backscattering method at future TeV linear colliders
one can obtain \GG\ and \GE\ colliding beams (photon colliders) with
the energy and luminosity comparable to that in \EPEM\ collisions. Now
this option is included to conceptual designs of linear colliders.
This paper (talk at the conference on nonlinear optics) is a short
introduction to this field with an emphasis on required lasers which
can be used both for e $\to \gamma$ conversion and for preparation of
electron beams (photoguns, laser cooling), some interesting nonlinear
QED effects in a strong field which apply restrictions on parameters
of photon colliders are discussed.
\end{abstract}

\section{INTRODUCTION}
  It is well known that according to Maxwell's equations
electromagnetic waves in  vacuum cross each other without
perturbations.  Quantum electrodynamics predicts nonzero cross section
of photon--photon scattering (via box diagram with virtual \EPEM\
pairs).  However for the optical photons this cross section is
extremely small, about $10^{-63}\;$cm$^2$  growing
very fast with the energy ($\propto \omega^6$ for $\omega \ll
m_{e}c^2$).  At high energies,photons can produced pairs of charged
particles (\EPEM, $\mu\mu, ... q\bar{q},$ WW etc.) and other final
states.

  Physics of elementary particles is studied in collisions of
particles of various types: $pp$, $p\bar{p}$, $ep$, \EPEM, etc..., which
allows to produce new particles, investigate their properties,
structure and interaction. Photon collisions are also of great
interest. Photons, as electrons, are elementary particles (that is
important for interpretation of results), cross sections of new
particle production in photon--photon collisions are of the same order
as that in \EPEM\ collisions. Many phenomena of particle physics can be
studied in \GG\ collision in the best way. Generally speaking, physics
in \GG\ collisions is complimentary to that studied in other types of
collisions. For example, \EPEM\  annihilation goes via one
virtual photon and resonances with odd charged parity (the same as for
photon) are produced: $\rho_0, \phi \ldots \Psi, \Upsilon$, Z.  While
in \GG\ collisions resonances with even charged parity are produced:
$\pi_0, \eta \ldots f, \eta_c, \chi \ldots$ H.

  Since 1970, two-photon physics has been actively studied at \EPEM\
storage rings. An electromagnetic field of a relativistic charged
particle is transverse and can be treated as equivalent photons. Approximately
accuracy one can assume that each electron is accompanied by almost
real photon with the probability d$n_{\gamma} \sim
(\alpha/\pi)\ln\gamma^2$ d$\omega/\omega \sim 0.035$ d$\omega/\omega$.
These experiments have given a lot of information on the nature of
elementary particles. For example, it was found that some of known
C-even mesons have very small cross sections in \GG\ collision that
indicates that they have not quark-antiquark nature (as usually) but
more likely they consist of four quark or two gluons.

However, the \GG\ luminosity (number of events/second is proportional to
luminosity) in collisions of virtual photons at storage rings is much
lower than \EPEM\ luminosity and concentrated mainly in the region of
low invariant masses of \GG\ system. In the region $W_{\GG}/2E_0 >0.5$
it accounts for only 0.02\% of \EPEM\ luminosity. Therefore, it is
natural that most of discoveries were done in \EPEM\ collisions.

  Maximum energy of \EPEM\ storage ring is about 100 GeV (LEP-II at
 CERN with circumference 27 km). Due to severe synchrotron radiation
 in storage rings the energy region beyond LEP-II can be explored only
 with linear colliders. Such linear colliders of 500 ---1500 GeV
 center-of-mass energy are developed now in the main high energy
 centers~\cite{LOEW}. Some parameters of the linear colliders are
 presented in table~\ref{tabl1}. Parameters of electron beams here are
 optimized for \EPEM\ collisions. The luminosity (L) of colliders
 given in the third line characterizes the rate of events and it is
 defined as follows: the number of events of a certain class per second
 $\dot{n} = L\sigma$, where $\sigma$ is the cross section of the
 process.  For cylindrical Gaussian beams $L \approx N^2
 f/4\pi\sigma_x\sigma_y $, where N is the number of particles in the
 beam, $\sigma_x, \sigma_y$ are transverse beam sizes, $f$ is the
 collision rate.

\begin{table}[t]
\caption{Some parameters of 0.5 TeV linear colliders, updated in 1996}
\begin{center}
\begin{tabular}{|l|c|c|c|c|c|c|} \hline
       project              &TESLA&SBLC&JLC(X)&NLC&CLIC&VLEPP\\ \hline
    country/center    &Germany&Germany&Japan&USA&CERN&Russia\\ \hline

$L,10^{33}\; \CMS $& 6 &  5.3  &  5.1  & 5.5 & 6.4 &  9.7  \\ \hline
rep.rate,Hz          & 5 & 50 & 150   & 180    & 700  &  300  \\ \hline
\# bunch/train      & 1130   & 333 & 85    & 90    & 20  &  1  \\ \hline
part./bunch($10^{10})$& 3.6  &1.1 & 0.65  & 0.75     & 0.8  &  20  \\ \hline
$\sigma_x(nm)$     &845 &335  & 260   & 295      & 264  & 2000  \\ \hline
$\sigma_y(nm)$     & 19   & 15   & 3     & 6.3       & 5   & 4    \\ \hline
$\sigma_z(mm)$     & 0.7& 0.3 & 0.09  & 0.125    & 0.16  & 0.75     \\ \hline
$\Delta$t bunch(ns)& 708  & 6 & 1.4   & 1.4    & 1   & ---  \\ \hline
\end{tabular}
\end{center}
\label{tabl1}
\end{table}

\begin{figure}[!htb]
\begin{minipage}[b]{0.45\linewidth}
\centering
\vspace*{0.5cm} 
\hspace*{1cm} \epsfig{file=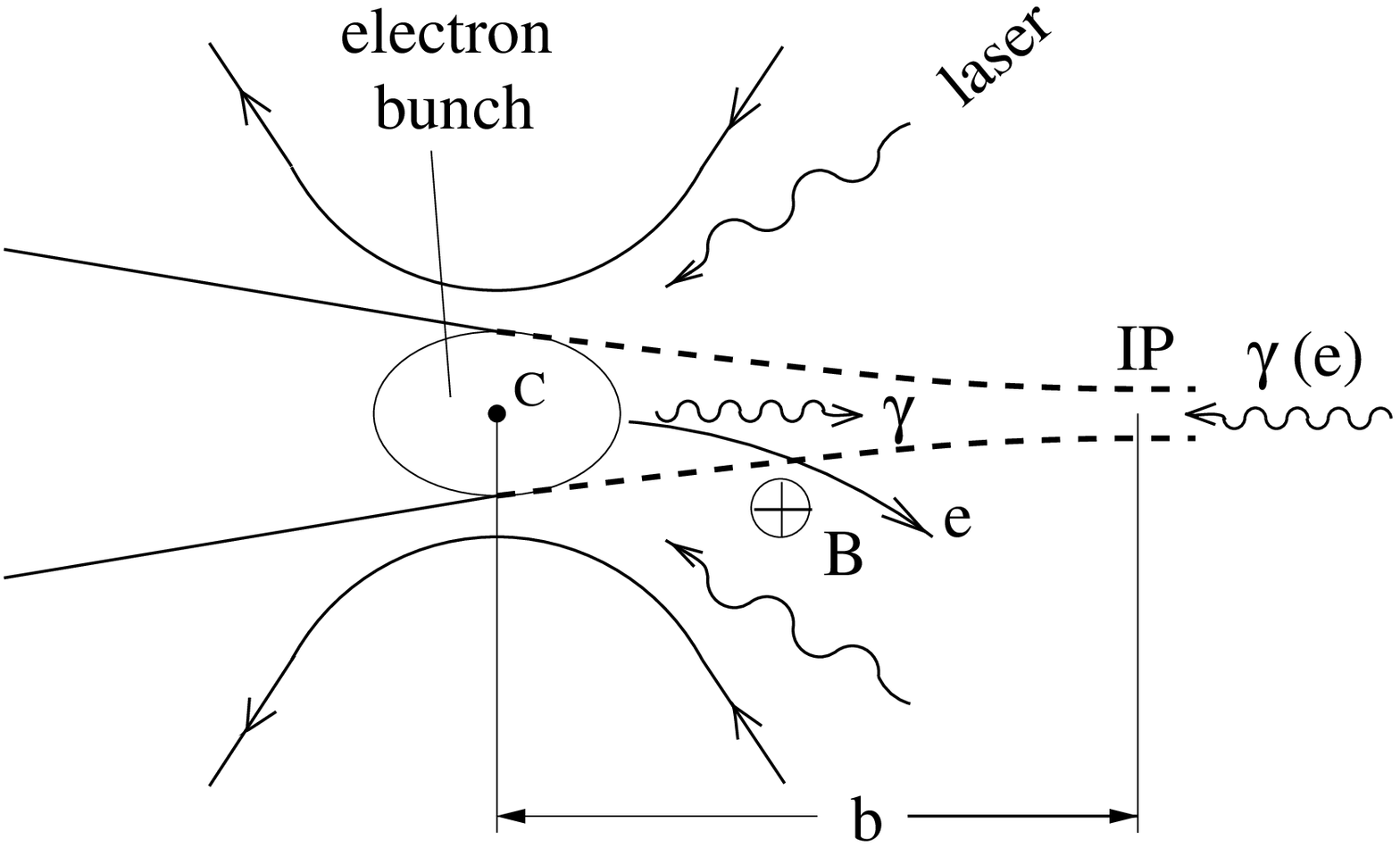,height=8.5cm,angle=-90} 
\vspace*{0cm}
\caption{Scheme of  \GG; \GE\ collider.}
\label{ris1}
\end{minipage}%
\hspace*{2cm} \begin{minipage}[b]{0.45\linewidth}
\centering
\vspace*{-0.5cm} 
\hspace*{0cm} \epsfig{file=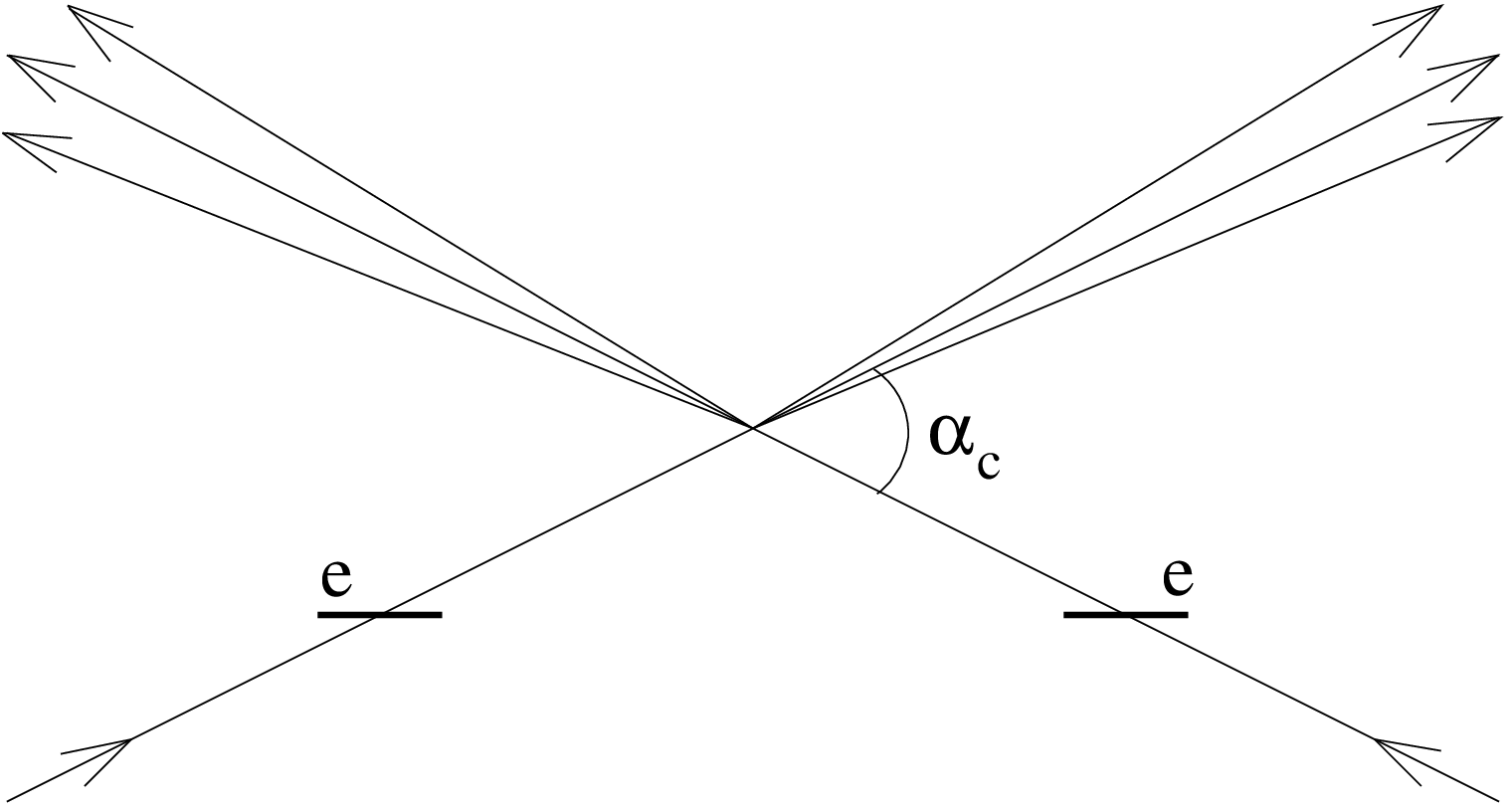,width=1.3in,angle=-90}
\vspace*{1.5cm}
\caption{Crab-crossing scheme}
\label{fig9}
\end{minipage}
\end{figure}

   Linear colliders offer unique, much more rich than before,
opportunities to study \GG\ and \GE\ interactions.  Unlike the
situation in storage rings, in linear colliders each bunch is used
only once.  This makes it possible to "convert"electron to high energy
photons to obtain colliding \GG\ and \GE\
beams~\cite{GKST81,GKST83}.  Among various method of $\gamma
\to$ e conversion the best one is Compton scattering of laser light on
high energy electrons.  The basic scheme of a photon collider is the
following, fig.\ref{ris1}.  Two electron beams after the final focus
system are traveling toward the interaction point (IP).  At a distance
of about 0.1--1 cm upstream from the IP, at the conversion point (C),
the laser beam is focused and Compton backscattered by electrons,
resulting in the high energy beam of photons.  With reasonable laser
parameters one can ``convert'' most of electrons into high energy
photons. The photon beam follows the original electron direction of
motion with a small angular spread of order $1/\gamma $, arriving at
the IP in a tight focus, where it collides with the similar opposing
high energy photon beam or with an electron beam. The photon spot size
at the IP may be almost equal to that of electrons at IP and
therefore, the luminosity of \GG, \GE\ collisions will be of the same
order of magnitude as the ``geometric'' luminosity of basic $ee$
beams.

Two collision scheme were considered. In the first scheme there is no
magnetic deflection of spent electrons and all particles after the
conversion region travel to the IP. The conversion point may be
situated very close to the IP. In the second scheme after the
conversion region particles pass through a region with a transverse
magnetic field ($B\sim$ 0.5--1 T) where electrons are swept
aside. Thereby one can achieve more or less pure \GG\ or \GE\
collisions.  

In both schemes the removal of the disrupted spent beams (with very
wide energy spread, $E=(0.02-1)E_0$, can best be done using the
crab-crossing scheme, fig.~\ref{fig9}. In this scheme the electron
bunches are tilted (using RF cavity) with respect to the direction of
the beam motion, and the luminosity is then the same as for head-on
collisions. Due to the collision angle $\alpha_c \sim 30$ mrad the
outgoing disrupted beams travel outside the final quads.

The detailed description of photon colliders properties can be found
in refs~\cite{GKST83}--\cite{TEL95} and in the Berkeley Workshop
Proceedings~\cite{BERK}.  Recently this option was included into the
Conceptual Design Reports of the NLC~\cite{NLC},
TESLA--SBLC~\cite{TESLA} and JLC. All these linear collider
projects foresee an interaction region for \GG,\GE\
collisions. 

  Physical problems, which can be studied in \GG\ and \GE\ collisions,
were discussed in  hundred papers.  Recent review of physics at
photon colliders can be found, for example, in TESLA/SBLC Conceptual Design
Report~\cite{TESLA}. 

 These studies of possible physics and technical problems have shown
that photon colliders are quite realistic and can substantially add to
a discovery potential of linear colliders. Below is some list of
arguments in favor of photon colliders.
\begin{itemize}
\item Some phenomena can be studied at photon colliders better than
  anywhere, for example: the measurement of \GG\ $\to$ Higgs 
  \footnote{Higgs boson (H) is undiscovered particle predicted in the
  unified theory of electroweak interactions, responsible for origin
  of particle masses. Its mass is not predicted in the theory, but
  from indirect experiments follows that 80 GeV $<M_Hc^2 <$ 400 GeV, that
  is accessible for $p\bar{p}$ storage rings and linear \EPEM,
  \EMEM, \GE, \GG\ colliders.} width which is sensitive to all heavy charged
  particles (even with $mc^2$ much larger than the collider energy);
  study of the vertex $\gamma\gamma$WW.

\item Cross sections for the pair production of charged scalar, leptons and
  top in \GG\ collisions are larger than those in \EPEM\ collisions by a
  factor of 5; for WW production this factor is even larger: 10--20.

\item In \GE\ collisions charged supersymmetric particles with masses
higher than in \EPEM\ collisions  can be produced
(heavy charged particle plus light neutral).

\item The luminosity of photon colliders (in the high energy part of
luminosity spectrum) with electron beam parameters considered in the
present designs will be about $10^{33}\; \CMS$ or by a factor 5 smaller
than $L_{\EPEM}$.  But the absence of collisions effects at 0.1 -- 1 TeV
photon colliders allows to reach \LGG\ up to $10^{35}$ \CMS\ using electron
beams with very low emittances. High luminosity photon colliders can
provide much higher production rate of WW pair and other charged
particles than in \EPEM\ collisions (see item 2) .

\item Obtaining of the ultimately high luminosities requires the
development of new techniques, such as the laser cooling of electron
beams~\cite{TSB1}. However, linear colliders will appear (may be) only
in one decade and will work next two decades.  The upgrading of the
luminosity requires the injection part modification only; it may be a
separate injector for a photon collider, merging of many low emittance
RF-photoguns (with or without laser cooling) is one of possible
variants. Note that for photon colliders positron beams are not
required that simplifies the task. 

\item Development of X-ray FEL lasers based on linear colliders (which
 are now under way) will favour the work on FEL required for photon
 colliders.

\item Summarizing we can say that physics goals for \GG,\GE\ colliders
are significant and complimentary to those in \EPEM\ collisions; the
technology (laser etc.) is sufficiently advanced and no more difficult
or risky than for \EPEM\ colliders, there are no show--stoppers;
incremental cost is relatively small;  therefore \GG,\GE\ colliders
should be build.
\end{itemize}

  Present paper is mainly focused on the application of lasers in this
project, also nonlinear QED effects important for photon colliders are
discussed.

\section{LASER PARAMETERS, CONVERSION EFFICIENCY}

The generation of high energy $\gamma$-quanta by the Compton scattering of the
laser light on relativistic electrons is  well known method~\cite{ARUT} and it
has been used in many laboratories. However, usually, the conversion
efficiency of electron to photons $k=N_{\gamma}/N_e$ is very small,
only about $10^{-7}$---$10^{-5}$. At linear
colliders, due to small bunch sizes one can focus the laser more tightly to the
electron beams and get $k\sim 1$ at rather moderate laser flash energy.

\begin{figure}[htb]
\begin{minipage}[b]{0.45\linewidth}
\centering
\vspace*{-1.cm} 
\hspace*{-1cm} \epsfig{file=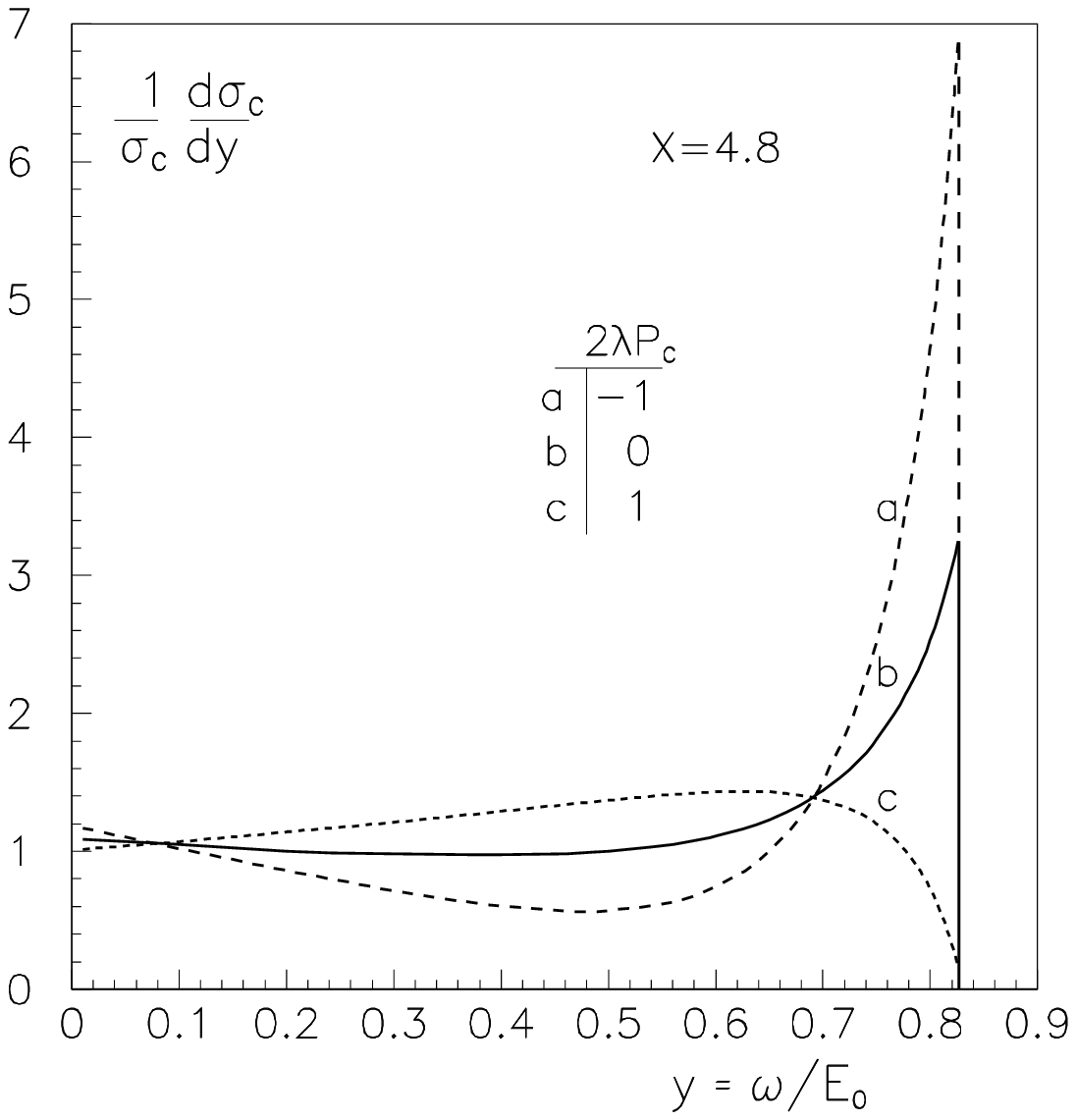, width=9cm}
\vspace*{-1.5cm}
\caption{ Spectrum of the Compton scattered photons for various
polarizations of laser and electron beams, see the text}
\label{fig4}
\end{minipage}%
\hspace*{1cm}\begin{minipage}[b]{0.45\linewidth}
\centering
\vspace*{-1.cm} 
\hspace*{-1.cm} \epsfig{file=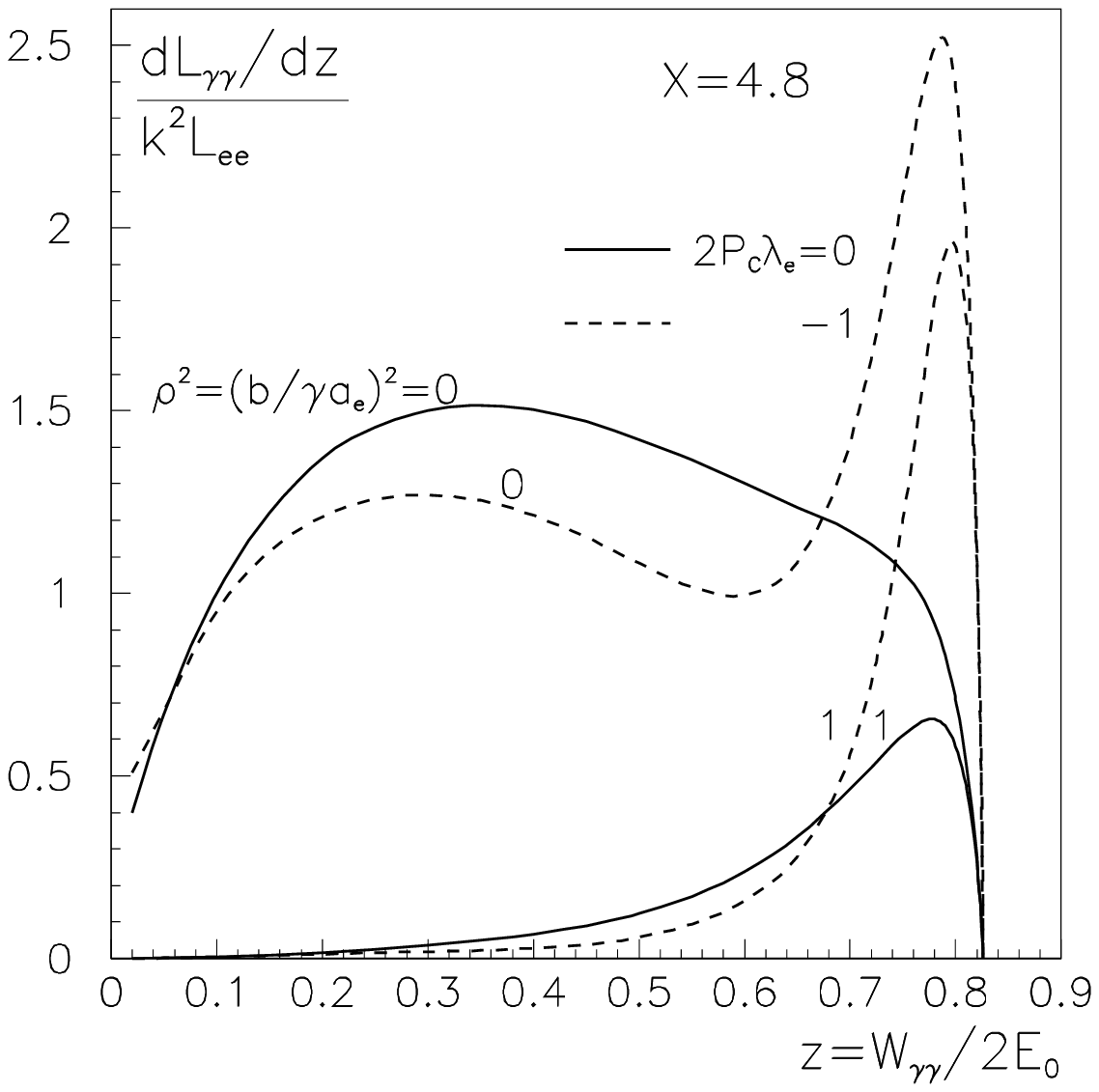, width=9cm}
\vspace*{-1.5cm}
\caption{Spectral luminosity of \GG\ collisions for various
polarizations and distances between conversion  and interaction
points, see the text}
\label{lum}
\end{minipage}
\end{figure}

In the conversion region, a laser photon with an energy $\omega_0$
collides almost head-on  with a high energy
electron of the energy $ E_0$.  The energy of the scattered photon
$\omega$ depends on its angle $\vartheta$ with respect to the motion of
the incident electron as follows~\cite{GKST83}:

\begin{equation}
\omega = \frac{\omega_m}{1+(\vartheta/\vartheta_0)^2},\;\;\;\;
\omega_m=\frac{x}{x+1}E_0; \;\;\;\;
\vartheta_0= \frac{mc^2}{E_0} \sqrt{x+1};
\end{equation}
where
\begin{equation}
x=\frac{4E \omega_0 }{m^2c^4}
 \simeq 15.3\left[\frac{E_0}{\TEV}\right]
\left[\frac{\omega_0}{eV}\right] = 
 19\left[\frac{E_0}{\TEV}\right]
\left[\frac{\mu m}{\lambda}\right],
\end{equation}
$\omega_m$ is the maximum energy of scattered photons.
For example: $E_0$ =250\,\, GeV, $\lambda=1.06$ \MKM, (Nd:Glass laser) 
$\Rightarrow$ x=4.5 and $\omega/E_0 = 0.82$.
The energy of the backscattered photons grows with
$x$. The spectrum also becomes narrower. However, at $x >
2(\sqrt{2}+1)\approx 4.8$ high energy photons are lost due to \EPEM\
creation in the collisions with laser
photons~\cite{GKST83,TEL90,TEL95}. For example, at
$x\sim10$, the maximum effective conversion coefficient is
$k_{max}\sim$ 0.3, while at $x < 4.8$ it can be about 0.65 (one
conversion length) or even high. The luminosity is proportional to
$k^2$ and in the latter case it is larger by a factor of 5. 
 So, the value $x \sim 4.8$ is optimum, though higher $x$ are also of
interest for the experiments where the ultimate monochromaticity of
\GG\ collisions is required.
 The wave length of the laser photons corresponding to $x=4.8$ is
\begin{equation}
 \lambda= 4.2 E_0 [\TEV]\;\; \mu m.
\end{equation}
For $2E_0 = 500\;$ \GEV\ it is about 1 \MKM, that is 
the region of the most powerful solid state lasers.
The energy spectrum of the scattered photons for $x=4.8$ is shown in
fig.~\ref{fig4} for various helicities of the electron and laser beams
(here $\lambda_e$ is the mean electron helicity ($|\lambda_e| \leq
1/2$), $P_c$ is the helicity of laser photons. We see that with the
polarized beams at $2\lambda_eP_c = -1$ the number of high energy
photons is nearly double. 
The spectrum in fig.~\ref{fig4} corresponds
to the case of small conversion efficiency. In the thick target each
electron may undergo the multiple Compton scattering. The secondary
photons are softer in general and populate the low energy part of the
spectrum.
If a laser light is polarized, high energy photons are also
polarized. Corresponding curves can be found
elsewhere~\cite{GKST84,TEL95}.  The polarization is very
advantageous for many experiments.

\IND\ The spectrum of scattered photons is very broad, but because of
energy-angle correlation in the Compton scattering it is possible to
obtain rather narrow distributions of the spectral luminosities in
\GG\ and \GE\ collisions.  Spectral luminosity distributions depend on
the variable $\rho=b/\gamma a$, where $b$ is the distance between the
conversion region and the interaction point (i.p.), $a$ is the
r.m.s. radius of the electron beam at the i.p.. If $\rho \ll 1$, then
at i.p.  photons of various energies collide with each other and the
distribution in the invariant mass of \GG\ or \GE\ system is
broad. But, if $\rho \gg 1$, then in \GE\ collisions electrons collide
only with the highest energy photons, therefore the invariant mass
spectrum of \GE\ collision is narrow.  In \GG\ collisions at $\rho \gg
1$ the photons with higher energy collide at smaller spot size and,
therefore, contribute more to the luminosity. As a result, the
luminosity spectrum is much narrower than at $\rho \ll 1$.  In
fig.~\ref{lum} the plots of spectral \GG\ luminosities are shown in
collisions of round beams at a low conversion coefficient.  The cases
of unpolarized and polarized beams($2P_c\lambda_e=-1$ both beams) for
$\rho= 0$ and $1$ are presented~\cite{TEL95}.  One can see that at
$\rho =1$ the luminosity in the low mass region is strongly suppressed
and the full width at half of maximum is about 10\% for polarized and
20\% for unpolarized beams. In realistic cases one should take into
account photons from the multiple Compton scattering and emitted by
electrons in the strong field of the opposing beam (so called
``beamstrahlung''), examples of realistic luminosity distribution can
be found elsewhere~\cite{TEL95,TESLA}.

A laser flash energy required for obtaining conversion efficiency
close to 100\% can easily be estimated using well known
characteristics of a laser beam with diffraction divergence. The
emittance of the Gaussian laser bunch with diffraction divergence is
$\epsilon_{x,y} = \lambda/4\pi$. The r.m.s. spot size of a laser beam
at the focus ($i=x,y$)~\cite{GKST83},
$ \sigma_{L,i}(0) = \sqrt{\lambda Z_R/4\pi}$,
 where $Z_R$ is known as the Rayleigh length.
The r.m.s. transverse size of a laser near the conversion region
depends on a distance $z$ to the focus (along the beam) as 
$
\sigma_{L,x}(z)= \sigma_{L,x}(0) \sqrt{1+z^2/Z_R^2}.
$
  We see that $2Z_R$ is the effective length of the conversion region
 is about $2Z_R$.
Neglecting multiple scattering, the dependence of the conversion
coefficient on the laser flash energy  A can be written as
\begin{equation}
     k  = N_\gamma /N_e \sim 1-\exp (-A/A_0 ),
\end{equation}
where $A_0$ is the laser flash energy for which the thickness of the 
laser target is equal to one Compton collision length.
This corresponds to $n_{\gamma}\sigma_{c}l = 1$, where
$ n_\gamma \sim A_0/(\pi \omega_0 a_{\gamma}^{2} l_\gamma )$,
$\sigma_c$ - is the Compton cross section ($\sigma_c =
1.8\times10^{-25}\;$ cm$^2$ at $x=4.8$), $l$ is the length of the region
with  high photon density,
which is equal to $2Z_R$ at $Z_R \ll \sigma_{L,z}\sim\sigma_z$
($\sigma_z$ is the electron bunch length). This gives
\begin{equation}
  A_0 = \frac{\pi\hbar c\sigma_z}{\sigma_c} = 5 \sigma_z [\MM],\,\,J
  \;\;\; \mbox{for}\;\; x=4.8.
\end{equation}
Note, that the required flash energy decreases with reducing the
Rayleigh length down to $\sigma_z$, and it is hardly changed with
further decrease in $Z_R$. This is because the density of photons
grows but the length having a high density decreases and the Compton
scattering probability remains almost the same.  It is not helpful to
make the radius of the laser beam at the focus smaller than
$\sigma_{L,x} \sim \sqrt{\lambda\sigma_z/4\pi}$, which may be much
larger than the transverse electron bunch size at the conversion
region.

Above we have considered only the
geometrical properties of the laser beam and the pure Compton effect.
However, in the strong electromagnetic field at the laser focus,
multiphoton effects (non-linear QED) are important. 
Nonlinear effects are described by the parameter~\cite{LANDAU}
\begin{equation}
\xi^2 = (eB\hbar/m\omega_0 c)^2,
\end{equation}
where $B$ is the r.m.s. strength of the magnetic field in the laser
wave. At $\xi^2 \ll 1$ an electron interacts with one photon (Compton
scattering), while at $\xi^2 \gg 1$ an electron scatters on many laser
photons simultaneously (synchrotron radiation in a wiggler).  The
transverse motion of an electron in the electromagnetic wave leads to
an effective increase in the electron mass: $m^2 \rightarrow
m^2(1+\xi^2)$, and the maximum energy of scattered photons decreases:
$\omega_m = x/(1+x+\xi^2)$.  At $x=4.8$, the value of $\omega_m/E_0$
decreases by 10\% at $\xi^2=0.6$.  Although this can be compensated by
some increase in $x$, but, in any case, large $\xi^2$ leads to
smearing of the high energy Compton peak.  The value of $\xi^2 \sim
1$ can be considered as the limit.

The nonlinear QED effects are more important for projects with shorter
bunches and higher beam energies (larger $\lambda$) and makes problems
already for some current projects~\cite{TEL90,TEL95}. In principle, in
order to avoid this problem one can make the laser target of less dense
but longer (to keep the conversion coefficient constant).  But in this
case, the required flash energy should be larger than that deduced
from the diffraction consideration only.

Recently~\cite{TSB1}, it was found how to avoid this problem. In the
suggested scheme the focus depth is stretched without changing the
radius of this area. In this case, the collision probability remains
the same but the maximum value of the field is smaller.  The solution
is based on use of chirped laser pulses~\cite{STRIC}  (wave
length is linearly depends on longitudinal position) and chromaticity
of the focusing system.  In the proposed scheme, the laser target
consists of many laser focal points (continuously) and light comes to
each point exactly at the moment when the electron bunch is there.
One can consider that a short electron bunch collides on its way
sequentially with many short light pulses of length $l_{\gamma} \sim
l_e$ and focused with $2Z_R \sim l_e$.  The required flash energy in
the scheme with a stretched laser focus is determined only by
diffraction and at the optimum wave length does not depend on the
collider energy.

\section{LASERS}

For conversion of electrons to photons with
$k\sim65$\% at $x=4.8$ and $E_0=250$~GeV a laser with the following 
parameters is required:

\vspace*{3mm}
\begin{tabular}{llll}
Power & $P\sim0.7$~TW & Duration & $\tau(rms)\sim\sigma_z/c\sim 1 - 2.5$~ps \\
Flash energy & $1 - 4$ J & Rep. rate & $\sim 10^4$ \\
Average power & $\sim 25$~kW & Wave length & $\lambda=4.2E0$[TeV], $\mu$m
\end{tabular}
\vspace*{3mm}

\noindent Obtaining of such parameters is possible with  solid
state or free electron lasers (FEL). For $\lambda\ge1\mu$m
($E_0 > 250$~GeV) FEL is the only option seen now.

\subsection{Solid state lasers}

In the last ten years the technique of short powerful lasers made an
impressive step and has reached petowatt ($10^{15}$) power levels and
few femtosecond durations~\cite{PERRY}.  Obtaining few joule pulses of
picosecond duration is not a problem. For photon collider applications
the main problem is a high repetition rate.  This is connected with
overheating of the amplifying media.

  The success in obtaining picosecond pulses is connected with a
chirped pulse amplification (CPA) technique~\cite{STRIC}.  The
principle of CPA is the following.  A short, low energy pulse is
generated in an oscillator.  Then this pulse is stretched by a factor
about $10^4$ in the grating pair which has delay proportional to the
frequency. This long nanosecond pulse is amplified and compressed by
another grating pair to a pulse with the initial or somewhat longer
duration. Due to practical absence of non-linear effects for the
stretched pulses, the obtained pulses have a very good quality close
to the diffraction limit.

   One of such lasers~\cite{MEYER95} works now in the E-144 experiment
studying nonlinear QED effects in the collision of laser photons
and 50 GeV electrons. It has a repetition rate of 0.5 Hz, 
$\lambda=1.06\;\mu$m (Nd:Glass), 2J flash energy, 2 TW power and
1 ps duration. This is a top-table laser with flash lamp pumping. 
Its parameters are 
very close to our needs, only the repetition rate is too low.
 Further progress in the
repetition rate (by two orders) is possible with a diode pumping (high
efficiency semiconductor lasers).  With diode pumping the efficiency
of solid state lasers reaches a 10\% level. Recent
studies~\cite{MEYER95,CLAY95,NLC} have shown that the combination of
CPA, diode pumping, zig-zag  slab amplifier, recombining of
several lasers (using polarizers and Pockel cells or slightly
different wave lengths), and  other laser techniques (if necessary)
such as phase-conjugated mirrors, moving amplifiers allows already now
to build a solid state laser system for a photon collider.
     All necessary technologies are developing actively now
for other  applications. 
  
\subsection{Free Electron Lasers}

   Free electron lasers (FEL) are very attractive for gamma-gamma
collider. Indeed, FEL radiation is tunable and has always minimal
(i.e.  diffraction) dispersion.  The FEL radiation is completely
polarized: circularly or linearly for the case of helical or planar
undulator, respectively.  The problem of synchronization of the laser
and electron bunches at the conversion region is solved by means of
conventional  methods used in accelerator techniques. A FEL amplifier
has potential to provide high conversion efficiency of the kinetic
energy of an electron beam into coherent radiation, up to 10$\%$. At
sufficient peak power of the driving electron beam the peak power of
the FEL radiation could reach the required TW level.
A short review of activities on FEL for photon colliders and references
can be found elsewhere~\cite{TESLA}.

\section{Optics at the interaction region}
\begin{figure}[thb]
\centering
\epsfig{file=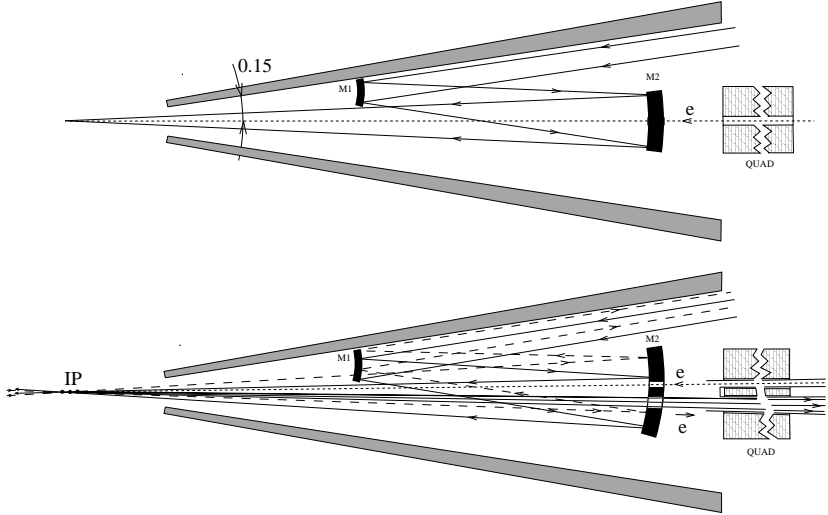,width=4.3in}
\caption{Layout of laser optics near the IP; upper - side view,
down - top view, dashed lines -- exit path of light coming from the
left through one of the CP points (right to the IP), the distance
between the IP and quads is about 2 m}
\label{fig21}
\end{figure}

 The possible layout of optics near the IP is shown in
 fig.~\ref{fig21}~\cite{TESLA}. The conversion region is situated at a
 distance $b \sim 1.5\gamma\sigma_y =$ 0.5--1.5 cm from the IP. Beams
 collideat the IP at the crab-crossing angle 30 mrad (see
 fig.~\ref{fig9}). The distance between IP and final quad is about 2
 m. All the optics is situated inside a tungsten conical mask which absorbs
 particle from showers on mirrors and quads. Arguments for chosing
 this geometry and estimation of mirror resistance are given in
 ref.~\cite{TESLA}.

\section{LASER COOLING}

To attain high luminosity, beams in linear colliders should be very
tiny. At the interaction point (IP) in the current LC
designs~\cite{LOEW}, beams with transverse sizes as low as \SX/\SY\
$\sim$ 200/4 nm are planned. Beams for \EPEM\ collisions should be
flat in order to reduce beamstrahlung energy loss (synchrotron
radiation in the field of the opposing beam). For \GG\ collision, the
beamstrahlung radiation is absent and beams with smaller \SX\ can be
used to obtain higher luminosity.

The transverse beam sizes are determined by the emittances \EX, and
\EY.  The beam sizes at the interaction point (IP) are
$\sigma_i=\sqrt{\EI\ \BI\ }$, where \BI\ is the beta function at the
IP (analog of the Rayleigh length in optics) which should be of the
order of the electron bunch length.  With the increase in the beam
energy the emittance of the bunch decreases: $\EI=\ENI/\gamma$, where
$\gamma=E/mc^2, $ \ENI\ is the {\it normalized} emittance.

The beams with a small \ENI\ are usually prepared in damping rings
which naturally produce bunches with $\ENY\ll\ENX$.  Laser
RF photoguns can also produce beams with low emittances.
However, for linear colliders it is desirable to have smaller emittances.

  Recently~\cite{TSB1}, a new method was considered --- laser cooling
of electron beams --- which allows, in principle, to obtain the
electron beam with cross sections by about two orders smaller than
attainable with other methods.

The idea of laser cooling of electron beams is very simple.  During a
collision with optical laser photons (in the case of strong field it
is more appropriate to consider the interaction of an electron with an
electromagnetic wave) the transverse distribution of electrons
($\sigma_i$) remains almost the same. Also, the angular spread
($\sigma_i^{\prime}$) is almost constant, because for
photon energies (a few eV) much lower than the electron beam energy
(several GeV) the scattered photons follow the initial electron
trajectory with a small additional spread. So, the emittance $\EI =
\SI \SIP$ remains almost unchanged. At the same time, the electron
energy decreases from $E_0$ down to $E$. This means that the
transverse normalized emittances have decreased: $ \EN = \gamma \E =
\EN_0(E/E_0)$.  One can reaccelerate the electron beam up to the
initial energy and repeat the procedure. Then after N stages of
cooling $ \EN /\EN _0 = (E/E_0)^N$ (if \EN\ is far from its limit).

 Some possible sets of parameters for the laser cooling are: $E_0 =
4.5$ GeV, $l_e=0.2 $ mm, $\lambda = 0.5$ \MKM, flash energy $A \sim 10
$ J.  The final electron bunch will have an energy of 0.45 \GEV\ with
an energy spread $\sigma_E/E \sim 13 \%$, the normalized emittances
\ENX,\ENY\ are reduced by a factor 10.  A two stage system with the
same parameters gives 100 times reduction of emittances. The limit on
the final emittance is $\ENX\ \sim\ENY\ \sim2\times 10^{-9}\;$ m~rad.
For comparison, in the TESLA (NLC) project the damping rings have
$\ENX\ =14(3)\times 10^{-6}\;$ m~rad, $\ENY\ =25(3)\times 10^{-8}\;$
m~rad. The electron beam after laser cooling accelerated to one TeV
energy can be focused to a spot with  about 0.5 nm diameter.

  For obtaining  ultimate emittances the parameter $\xi^2$ in the
  cooling region should be less than one (Compton
  scattering)~\cite{TSB1}. This is also necessary  for preserving
  the electron polarization. These conditions can be achieved using
  the method of laser focus stretching  described in sect.2. 

Laser cooling requires a laser system even more powerful than that for e $\to
\gamma$ conversion. However, all the requirements are reasonable taking
into account fast progress of laser technique and time plans of linear
colliders. A multiple use of the laser bunch can reduce considerably an average
laser power.

   Using laser cooling one can attain in \GG\ collisions the
luminosity by one order higher than that in \EPEM\
collisions~\cite{TSB2,PH97}. Besides, the typical cross section in the
\GG\ collisions are larger by a factor of five. Both these facts are
very good arguments in favor of photon colliders.

\section{MAXIMUM ENERGY AND LUMINOSITY OF PHOTON COLLIDERS}

  Cross sections of ``interesting'' physical processes (such as charged
pair production) decrease usually with  energy as $1/E^2_{cm}$, where
$E_{cm}$ is the center-of-mass energy of colliding particles, therefore the 
luminosity of colliders should grow proportionally to $E_{cm}^2$. A reasonable
scaling for the required \GG\ luminosity at \GG\ collider is
\begin{equation}
\LGG\ \sim 3\times10^{33}E^2_{cm}\;\;\CMS,
\label{scal}
\end{equation}
where $E_{cm}$ is in TeV unit. In \EPEM\ collisions characteristic
cross sections are somewhat smaller and a required luminosity is
larger by a factor of 5. 

  In linear colliders, each bunch is used only once, its production and
acceleration requires a certain power. It means that the number of
accelerated particle can not be increased with  energy. In current
projects a power consumption is already 100--200 MW. The only way to
increase the luminosity is a decrease in transverse beam sizes at the
interaction point.

In \EPEM\ collision, the  minimum beam sizes are determined by strong
radiation of particles in the field of the opposing beam
(beamstrahlung) and beam--beam instability. These effects impose
strong restrictions on beam parameters and determine the attainable
luminosity in \EPEM\ collisions. In table 1 one can see that all
projects use the flat beams, vertical size is of the order of 3--20 nm and
horizontal size is larger by a factor of 100. This is because for the
same beam cross section  the electromagnetic field is reverse
proportional to the largest transverse size.

\subsection{Restrictions due to coherent pair creation}

  At the first sight, in \GG\ collisions both beams are neutral and
collision effects should be absent. It is not a case. The conversion point
is situated very close to the interaction point (IP), and electrons
after conversion pass the IP very close to the axis and influence on the
oncoming photons and electrons. 

What happens with photons?  First of all, a photon can be converted into
\EPEM\ pair in a collision with an individual electron (Bethe-Heitler
process --- incoherent pair creation). The cross section of this
process at  high energies is about $4\times 10^{-26}\;$cm$^2$. The
characteristic luminosity in one bunch collision is about $10^{30}$
\CMS\ therefore as much as $10^4$ \EPEM\ pairs will be produced per one bunch
collisions (these pairs move in forward direction and do not cause
serious background in detector). Note, that this process does not
destroy the beam containing about $10^{10}$ photons.

More important for photon colliders is the process of {\it coherent
pair creation} (conversion of a photon into \EPEM\ pair in the
collective field of an opposing electron
beam)~\cite{CHEN,TEL90,TEL95,TSB2}.  This process becomes important
for the beam fields $B$ ($|E|+|B|$ in our problem) when $\kappa =
(\omega/mc^2)/(B/B_0) >1,\;\; $ where $ B_0= m^2c^3/e\hbar = \alpha
e/r_e^2 = 4.4 \times 10^{13}\;$ G is the critical field, $r_e =
e^2/mc^2$ is the classical radius of electron, $\omega$ is the photon
energy. The origin of this condition is the following. In the rest
system of the virtual \EPEM\ pair created by the photon the electrical
field of the beam is $E^{\prime}= (\omega/mc^2)\times B$. If the
energy acquired by $e^+$ and $e^-$ in this field on the Compton wave
length $\lambda_c=\hbar/mc$ is larger than $mc^2$ then a virtual
particle may become a real one. That is a spontaneous pair creation in
a strong field. In our problem the field of the opposing beam in the
reference system of the oncoming beam can exceed the critical field
$B_0 = 4.4\times10^{13}$ G. So, the critical field can be exceeded not
only in stars but also at high energy linear colliders.

  The coherent pair creation can impose restriction on the value of
  attainable luminosity. Fortunately, at an energy below about one
  TeV (which is a goal of the next linear colliders) this effect is
  considerably suppressed due to repulsion of  electron beams during
  collisions. Even infinitely narrow electron beams after
  repulsion at the IP create an acceptable field on the axis, which
  does not cause catastrophic loss of the \GG\ luminosity.

\begin{figure}[!htb]
\centering
\vspace*{1cm}
\epsfig{file=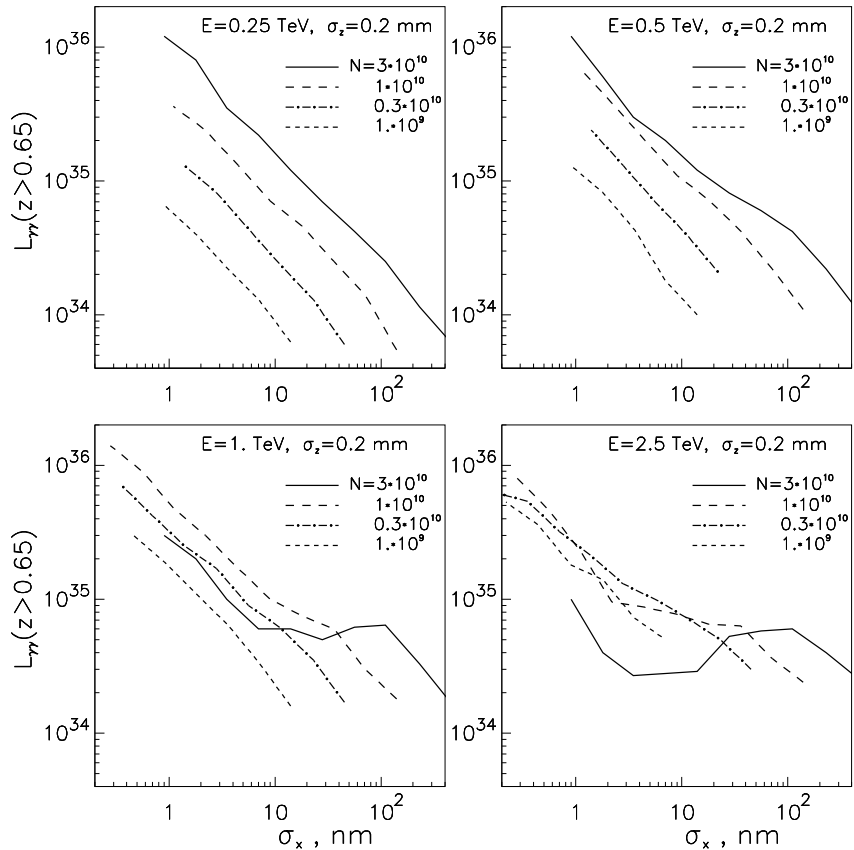,width=11.5cm,bb= 32 50 550 540 }
\caption{Dependence of the \GG\ luminosity on the horizontal beam size
for $\sigma_z = 0.2$ \MM, see comments in the text.}
\label{sb3}
\end{figure}

 Some simulation results  of the attainable \GG\
 luminosity are shown in fig.~\ref{sb3}\cite{TSB2}. It was
 assumed that the conversion point is situated as close as possible to
 the IP at a distance $b= 3\sigma_z + 0.04E[\TEV]$ cm. The second
 number is equal to the minimum length of the conversion region posed by
 nonlinear QED effects. The vertical electron beam size was taken
 smaller than $b/\gamma$, the horizontal size was varied. The total
 beam power was kept equal to $15E_{cm}[\TEV]$ MW.
In fig.~\ref{sb3} one can see that, if the collision effects are not
 important, the luminosity curves follow their natural behavior $L
 \propto 1/\sigma_x$. This is valid for all curves at large horizontal
 beam sizes (low beam fields). At low $\sigma_x$ some curves make zigzag or
change slope, this is due to the conversion of photons into \EPEM\ pairs. 

These results lead us to following conclusions. For the energies $E_{cm}
<$ 2 TeV, which are in reach of the next generation of linear colliders, the
luminosity limit is much higher than that required by our scaling for
luminosity given by eq.~\ref{scal}. The energy limit for photon
colliders is about $E_{cm}\sim$ 5--10 TeV.

\subsection{Depolarization of photons in a strong beam field}.

It is well known that the region with a strong electromagnetic field can
be treated as an anisotropic medium with some refraction
index~\cite{LANDAU}. In fact, the conversion of photons to \EPEM\ pairs
(absorption) considered in the previous section is the  manifestation of
the imaginary part of the refraction index.  The values  $n$ are
different for photons with linear polarization parallel ($n_1$) and
perpendicular ($n_2$) to the field direction. Similar to the 
behaviour of optical photons in anisotropic crystals an electromagnetic field
in the vacuum can change a photon polarization. Such an effect was
considered before for high energy photons traveling in a strong field
between crystal planes~\cite{BAIER2}. A similar effect (but not in
constant field) is possible in the conversion region at photon
colliders where the high energy photon polarization can change due to
interaction with high density optical photons in the laser
focus~\cite{SERBO}. Here we will consider the new effect at photon
colliders -- an influence of the field of the opposite electron beam on the
polarization of high energy photons. A detailed consideration of this
effect will be published elsewhere~\cite{TELPOL}, here only main
results are presented.

 The field of a relativistic beam has the electrical ($E_b$) and magnetic
 ($B_b \approx E_b$) components perpendicular to each other and to the
 direction of beam motion. For the photon moving oppositely to this
 beam the effective field is $|B|=|E_b|+|B_e|$. Let us denote $n_1$ to be
 a refraction index for linear polarization of the photon (electrical
 field) parallel to the beam electrical field and $n_2$ for
 perpendicular orientation. 

\begin{figure}[!htb]
\centering
\vspace*{-1.3cm}
\epsfig{file=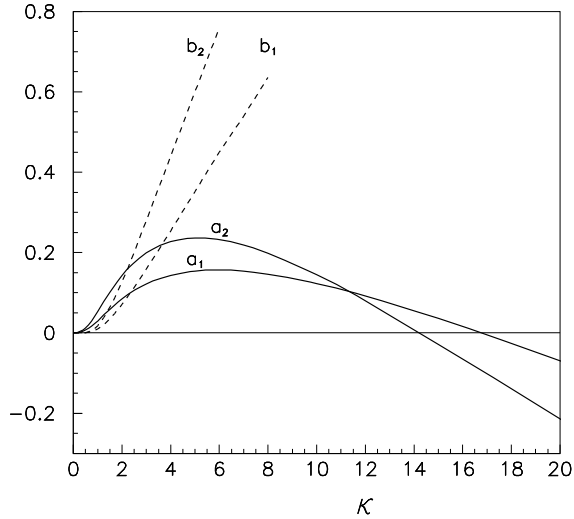,width=10cm}
\vspace*{-1.2cm}
\caption{Real and imaginary parts of a refraction index of the vacuum
with a strong and constant electromagnetic field:
$n^2-1=\alpha\,(a+ib)/\gamma^2, $where $\gamma=\hbar\omega/mc^2$; see
comments in the text.}
\label{nvac}
\end{figure}

Formula (rather complicated) for $n_1, n_2$ can be found
elsewhere~\cite{NAROZ,RITUS,BAIER1}. It can be written in the form
\begin{equation}
n_i^2-1=\alpha(a_i+i b_i)/\gamma^2,
\label{nab}
\end{equation}
where $\alpha=e^2/\hbar c=1/137,\;\gamma=\hbar \omega/mc^2,\; a_i$ and
$b_i \;(i=1,2)$ are the functions of the parameter
\begin{equation}
\kappa = \gamma \frac{B}{B_0}, 
\end{equation}
where $B_0=m_e^2c^3/e\hbar=\alpha e/r_e^2=4.4\times10^{13}$ G is
the critical field. Values of $a_i$ and $ b_i$ are plotted in fig.~\ref{nvac}.
Asymptotic values of $n_i$ are the following
\begin{eqnarray}
n_{1,2}-1 \approx &\frac{\alpha}{\gamma^2}\left(\frac{11\mp3}{180\pi}\kappa^2+
i\sqrt{\frac{3}{2}}\frac{3\mp 1}{32}\kappa
e^{-8/3\kappa}\right),\;&\kappa \ll 1 \nonumber \\
&\frac{\alpha}{\gamma^2}\left(-\frac{5\mp
1}{56\pi^2}\sqrt{3}\Gamma^4(2/3)(1-i\sqrt{3}) (3\kappa)^{2/3}\right),
\;&\kappa \gg 1
\end{eqnarray}

At $\kappa \leq 2$ the refraction is determined by the real part of
$n$, for larger $\kappa$ an imaginary part of $n$ connected with
absorption of photons is more important. Photons are absorbed due to
\EPEM\ pair creation. The probability of pair creation is connected
with $\Im n$.  For a circularly polarized photon the probability of
absorption on a collision length approximately equal to
r.m.s. bunch length ($\sigma_z$) is
\begin{equation}
W_{\EPEM} \approx \alpha^2\bar{b}(\kappa)\sigma_z/\gamma r_e.
\label{W}
\end{equation}  

\begin{figure}[!htb]
\centering
\vspace*{-1.3cm}
\epsfig{file=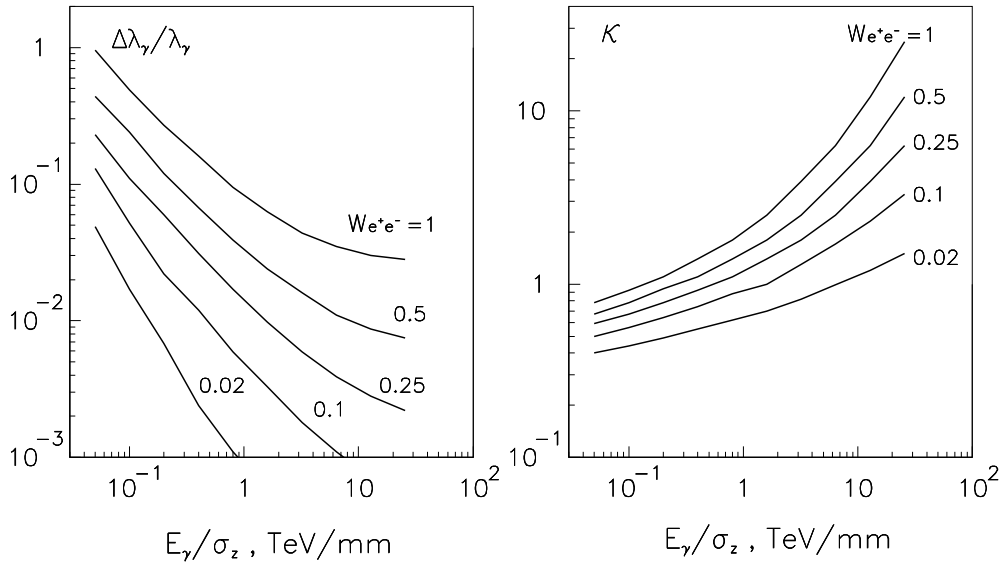,width=15cm}
\caption{ a) Decrease in a photon helicity during beam collisions for
various beam parameters and probabilities of coherent pair creation
(W); b) the parameter $\kappa = \gamma B/B_0$ for the conditions
corresponding to the left figure. The luminosity at fixed value of
abscissa is approximately proportional to $\kappa\times e^{-W}$, see
additional comments in the text.}
\label{depol}
\end{figure}

 We see that the beam field is an isotropic medium with different
 refraction indeces for linear polarization parallel and perpendicular
 to the beam direction. Such a medium transforms the circular $\gamma$
 quanta polarization of into the linear polarization and vice
 versa. In addition, due to the difference in the absorption length the total
 polarization decreases. Using eqs.~\ref{nab},\ref{W} one can obtain
 the decrease in circular polarization $\lambda_{\gamma}$ during the
 beam collision (it is assumed that the initial polarization was 100\%)
\begin{equation}
\frac{\Delta
\lambda_{\gamma}}{\lambda_{\gamma}}=\frac{\alpha^4}{8\gamma^2}
\left(\frac{\sigma_z}{r_e}\right)^2
(\Delta a^2 + \Delta b^2)=\frac{\Delta a^2 +\Delta
b^2}{8\bar{b}^2}W^2_{\EPEM},
\label{dP}
\end{equation}
here $\Delta a = a_2-a_1, \Delta b=b_2-b_1$, $a$ and $b$ are functions
of $\kappa$. By excluding (numerically) $\kappa$ from
eqs.~\ref{W},\ref{dP} we can find
$\Delta\lambda_{\gamma}/\lambda_{\gamma}$ as a function of
$\gamma/\sigma_z$ for different $W_{\EPEM}$, it is shown in
fig.~\ref{depol}a. For low $\gamma/\sigma_z$ and large  $W_{\EPEM}$ the
depolarization is unacceptably high.

The \EPEM\ pair creation is the only one effect restricting the \GG\
luminosity. In principle, it may happen that photons lose their
polarization at the luminosity much lower than its limit posed by the
pair creation. To check this we plotted in fig.~\ref{depol}b the values
of $\kappa$ as a function of the same parameters as those in
fig.~\ref{depol}a. In order to decrease $\Delta
\lambda_{\gamma}/\lambda_{\gamma}$ at fixed $\gamma/\sigma_z$ we have 
to decrease $W_{\EPEM}$ (see fig.~\ref{depol}a) which can be done, for
example, by increasing the horizontal beam size $\sigma_x$. This
corresponds to some decrease in $\kappa$ (see fig.~\ref{depol}b). Now
we can estimate the luminosity loss  in this procedure. The luminosity is
$L\propto (1/\sigma_x\sigma_y)\times e^{-W}\propto \kappa\times
e^{-W}$. Looking at fig.~\ref{depol}b one can find that the retention of
depolarization below 1$\%$ is accompanied by a very small loss of
luminosity (compared to the maximum one), it is less than 20$\%$ in
the whole region of parameters. This is very good. Note, however, that
so simple adjustments of polarization loss by varying the horizontal
beam size $\sigma_x$ is not always possible. In the case where the initial
electron beams are very narrow and the pair creation is suppressed only
due to the beam repulsion (as it was discussed in the previous section)
we will get some value of $\kappa$ which is not controlled by
the focusing of the electron beams. If depolarization degree is not acceptable
it is necessary to adjust the number of particles in the beam and its
length.
 
  So, the analysis of depolarization phenomena at photon colliders
connected with the anisotropy of the refraction index in the collision
region with a strong electromagnetic field shows that this effect is
essential in some cases and should  be taken into account. By proper
measures one can keep the depolarization below few percents without
noticiable loss of the \GG\ luminosity.

\section{Conclusion}
  
   Future linear colliders offer unique opportunities to study
particle physics in \GG\ and \GE\ interactions.  Such photon colliders
at an energy 0.1 -- 1 TeV are being developed now in the main high energy
centers. Hopefully, in 15--20 years this will be one of the main
areas of particle physics.

\end{document}